\documentclass[12pt]{article}
\newenvironment{eqnarraz}%
   {\setlength{\arraycolsep}{0.15em} \begin{eqnarray}}%
   {\end{eqnarray}}
\newenvironment{eqnarraz*}%
   {\setlength{\arraycolsep}{0.15em} \begin{eqnarray*}}%
   {\end{eqnarray*}}
   
\newtheorem{Lemma}{Lemma}

\newcommand{\eref}[1]{{(\ref{#1})}}

\newcommand{\rmd}{{\rm{d}}}
\newcommand{\rme}{{\rm{e}}}

\newcommand{\ket}[1]{ | #1  \rangle}

\newcommand{\rmi}{{\rm{i}}}
\newcommand{\ii}{{\rmi_1}}
\newcommand{\iii}{{\rmi_2}}
\newcommand{\jj}{{\rm{j}}}
\newcommand{\ee}{\mathbf{e_1}}
\newcommand{\eee}{\mathbf{e_2}}
\newcommand{\e}[1]{{\mathbf{e_{#1}}}}

\newcommand{\T}{\mathbb{T}}
\newcommand{\C}{\mathbb{C}}
\newcommand{\D}{\mathbb{D}}
\newcommand{\R}{\mathbb{R}}
\newcommand{\nc}{\mathcal{NC}}
\newcommand{\expm}[1]{\textrm{exp}\left\{ #1 \right\}}

\renewcommand{\P}[2]{P_{#1}( #2 )}

\renewcommand{\[}{\left[}

\usepackage{hyperref}
\usepackage{latexsym}
\usepackage{amsmath}
\usepackage{amssymb}

\begin{document}
\begin{center}
\LARGE
\textbf{The Bicomplex Quantum Harmonic
Oscillator}\\[1cm]
\large
\textbf{Rapha\"{e}l Gervais Lavoie$^1$, Louis Marchildon$^1$\\
and Dominic Rochon$^2$}\\[0.5cm]
\normalsize
$^1$D\'{e}partement de physique,
Universit\'{e} du Qu\'{e}bec,\\
Trois-Rivi\`{e}res, Qc.\ Canada G9A 5H7\\[0.5cm]
$^1$D\'{e}partement de math\'{e}matiques
et d'informatique,
Universit\'{e} du Qu\'{e}bec,\\
Trois-Rivi\`{e}res, Qc.\ Canada G9A 5H7\\[0.5cm]
email: raphael.gervaislavoie$\hspace{0.3em}a\hspace{-0.8em}\bigcirc$uqtr.ca,
louis.marchildon$\hspace{0.3em}a\hspace{-0.8em}\bigcirc$uqtr.ca,
dominic.rochon$\hspace{0.3em}a\hspace{-0.8em}\bigcirc$uqtr.ca\\
\end{center}
\medskip

\begin{abstract}
The problem of the quantum harmonic oscillator is
investigated in the framework of bicomplex numbers,
which are pairs of complex numbers making up a
commutative ring with zero divisors.
Starting with the commutator of the bicomplex position
and momentum operators, and adapting the algebraic
treatment of the standard quantum harmonic
oscillator, we find eigenvalues and eigenkets
of the bicomplex harmonic oscillator Hamiltonian.
We construct an infinite-dimensional bicomplex module
from these eigenkets.  Turning next to the
differential equation approach, we obtain
coordinate-basis eigenfunctions of the bicomplex harmonic
oscillator Hamiltonian in terms of hyperbolic Hermite
polynomials.
\end{abstract}
\section{Introduction}
\setcounter{equation}{0}
The mathematical structure of quantum mechanics
consists in Hilbert spaces defined over the field
of complex numbers~\cite{Neumann}.  This structure
has been extremely successful in explaining vast
amounts of experimental data pertaining largely,
but not exclusively, to the world of molecular,
atomic and subatomic phenomena.

That success has led a number of investigators,
over many decades, to look for general principles or
arguments that would lead quite inescapably to the
complex Hilbert space structure.  It has been
argued~\cite{stuec1,stuec2}, for instance, that the
formulation of an uncertainty principle, heavily
motivated by experiment, implies that a real Hilbert
space can in fact be endowed with a complex structure.
The proof, however, involves a number of additional
hypotheses that may not be so directly connected with
experiment.  In fact Reichenbach~\cite{reichenbach}
has shown that a theory is not straightforwardly
deduced from experiments, but rather arrived at by
a process involving a good deal of instinctive
inferences.  The justification of the theory lies in
its ability to explain known experimental results and
to predict new ones.  More recently, some of the efforts
to derive the complex Hilbert space structure have
focused on information-theoretic principles~\cite{Fuchs,Bub}.
The general principles assumed at the outset are no doubt
attractive, but yet open to questioning.  The upshot
is that there is no compelling argument restricting the
number system on which quantum mechanics is built to the
field of complex numbers.  A possible extension
of quantum mechanics to the field of quaternions was
pointed out long ago by Birkhoff and von
Neumann~\cite{Von Neumann}, and it has since been
developed substantially~\cite{Nash,Adler}.

The fields of real ($\mathbb{R}$) and complex
($\mathbb{C}$) numbers, together with the
(non-com\-mu\-ta\-tive) field
of quaternions ($\mathbb{H}$), share two properties
thought to be very important for
building a quantum mechanics.  Firstly, they are
the only associative division algebras over the
reals~\cite{Oneto}.  A \emph{division algebra} is
one that has no zero divisors, that is, no non-zero
elements $w$ and $w'$ such that $w w' = 0$.
Secondly, they are the only associative
absolute valued algebras with unit over the
reals~\cite{Albert}.  An \emph{absolute valued
algebra} is one that has a mapping $N(w)$ into
$\mathbb{R}$ that satisfies
\begin{enumerate}
\item $N(0) = 0$;
\item $N(w) > 0$ if $w \neq 0$;
\item $N(a w) = |a| N(w)$ if $a \in \mathbb{R}$;
\item $N(w_1 + w_2) \leq N(w_1) + N(w_2)$;
\item $N(w_1 w_2) = N(w_1) N(w_2)$.
\end{enumerate}
Property (v), in particular, is widely believed
crucial to represent quantum-mechanical
probabilities and the correspondence principle
with classical mechanics.

Yet several investigations have been carried
out on structures sharing some characteristics
of quantum mechanics and based on number
systems that are neither division nor
absolute valued algebras~\cite{Millard,Kocik}.
Of these number systems the ring $\mathbb{T}$ of
bicomplex numbers is among the simplest.
It has already been shown~\cite{Rochon3} that
structures analogous to bras, kets and
Hermitian operators can be defined in
finite-dimensional modules over~$\mathbb{T}$.

In this paper we intend to pursue that investigation
further by extending to bicomplex numbers the problem
of the quantum harmonic oscillator.  The harmonic
oscillator is one of the simplest
and, at the same time, one of the most important
systems of quantum mechanics, involving as it is
an infinite-dimensional vector space.

In section~2 we review the main properties of bicomplex
numbers that we will use, together with the notions
of module, scalar product and linear operator.
Section~3 is devoted to the determination of
eigenvalues and eigenkets of the bicomplex
quantum harmonic oscillator Hamiltonian, along
lines very similar to the algebraic treatment
of the usual quantum-mechanical problem.  To our
knowledge, this is the first time that such eigenvalues
and eigenkets are obtained with a number system
larger than~$\mathbb{C}$.  An
infinite-dimensional module over~$\mathbb{T}$ is
explicitly constructed with eigenkets as basis.
Section~4 develops the coordinate-basis
eigenfunctions associated with the eigenkets
obtained.  This leads to a straightforward and
rather elegant generalization of the usual
Hermite polynomials as hyperbolic functions of
a real variable.  Section~5 connects with standard
quantum mechanics and opens up on new problems.

\section{Bicomplex numbers and modules}\label{Preliminaries}
\setcounter{equation}{0}
This section summarizes basic properties of bicomplex
numbers and finite-dimensional modules defined
over them.  The notions of scalar product and
linear operators are also introduced.  Proofs
and additional material can be found
in~\cite{Rochon3,Price,Rochon1,Rochon2}.

\subsection{Algebraic properties of bicomplex numbers}
\label{Definition of bicomplex numbers}
The set $\mathbb{T}$ of \emph{bicomplex numbers}
is defined as
\begin{equation}
\mathbb{T}:=\{ w = w_e + w_{\ii} \ii + w_{\iii} \iii + w_j \jj
~|~ w_e, w_{\ii}, w_{\iii}, w_j \in \mathbb{R} \} ,\label{2.1a}
\end{equation}
where $\ii$, $\iii$ and $\jj$ are imaginary and hyperbolic
units such that $\rmi_1^2= -1 = \rmi_2^2$ and $\jj^2=1$.
The product of units is commutative and defined as
\begin{equation}
\ii\iii = \jj, \qquad \ii\jj = -\iii,
\qquad \iii\jj = -\ii. \label{2.2}
\end{equation}
With the addition and multiplication of two
bicomplex numbers defined in the obvious way,
the set $\mathbb{T}$ makes up a commutative ring.

Three important subsets of $\mathbb{T}$ can be
specified as
{\begin{eqnarraz}
\mathbb{C}(\rmi_k) &:=& \{ x+y\rmi_k~|~x,y\in\mathbb{R} \},
\quad k=1,2 ;\\
\mathbb{D} &:=& \{ x+y\jj~|~x,y\in\mathbb{R} \} .
\end{eqnarraz}}%
Each of the sets $\mathbb{C}(\rmi_k)$ is isomorphic
to the field of complex numbers, and $\mathbb{D}$ is
the set of \emph{hyperbolic numbers}.  An arbitrary
bicomplex number $w$ can be written as
$w = z + z' \iii$, where $z = w_e + w_{\ii} \ii$ and
$z' = w_{\iii} + w_j \ii$ both belong to $\mathbb{C}(\ii)$.

Bicomplex algebra is considerably simplified by
the introduction of two bicomplex numbers $\ee$
and $\eee$ defined as
\begin{equation}
\ee:=\frac{1+\jj}{2},\qquad\eee:=\frac{1-\jj}{2}.
\label{2.9}
\end{equation}
One easily checks that
\begin{equation}
\mathbf{e}_{\mathbf{1}}^2=\ee,\qquad
\mathbf{e}_{\mathbf{2}}^2=\eee, \qquad
\ee + \eee=1, \qquad\ee\eee=0 . \label{2.10}
\end{equation}
Any bicomplex number $w$ can be written uniquely as
\begin{equation}
w = z_1 \ee + z_2 \eee , \label{2.10a}
\end{equation}
where $z_1$ and $z_2$ both belong to $\mathbb{C}(\ii)$.
Specifically,
\begin{equation}
z_1 = (w_e + w_j) + (w_{\ii} - w_{\iii}) \ii ,
\qquad z_2 = (w_e - w_j) + (w_{\ii} + w_{\iii}) \ii .
\end{equation}
The numbers $\ee$ and $\eee$ make up the so-called
\emph{idempotent basis} of the bicomplex numbers.  Note that
the last of~(\ref{2.10}) illustrates the fact that
$\mathbb{T}$ has zero divisors which, we recall,
are non-zero elements whose product is zero.

The product of two bicomplex numbers $w$ and $w'$
can be written in the idempotent basis as
\begin{equation}
w \cdot w' = (z_1 \ee + z_2 \eee)
\cdot (z_1' \ee + z_2' \eee)
= z_1 z_1' \ee + z_2 z_2' \eee .
\label{2.4cc}
\end{equation}
Since 1 is uniquely decomposed as $\ee + \eee$,
we can see that $w \cdot w' = 1$ if and only if
$z_1 z_1' = 1 = z_2 z_2'$.  Thus $w$ has an inverse
if and only if $z_1 \neq 0 \neq z_2$, and the
inverse $w^{-1}$ is then equal to
$z_1^{-1} \ee + z_2^{-1} \eee$.  A non-zero $w$ that
does not have an inverse has the property that
either $z_1 = 0$ or $z_2 = 0$, and such a $w$ is
a divisor of zero.  Zero divisors make up the
so-called null cone $\nc$.  That terminology comes
from the fact that when $w$ is written as $z + z' \iii$,
zero divisors are such that $z^2 + (z')^2 = 0$.

With $w$ written as in~(\ref{2.10a}), we define two
projection operators $P_1$ and $P_2$ so that
\begin{equation}
P_1 (w) = z_1 , \qquad P_2 (w) = z_2 .
\end{equation}
One can easily check that, for $k=1, 2$,
\begin{equation}
[P_k]^2=P_k, \qquad \ee P_1 + \eee P_2 ={\rm Id}
\end{equation}
and that, for any $s,t\in\T$,
\begin{equation}
\P{k}{s+t}=\P{k}{s}+\P{k}{t},
\qquad \P{k}{s\cdot t}=\P{k}{s}\cdot \P{k}{t} . \label{2.16}
\end{equation}

We define the conjugate $w^{\dagger}$ of the bicomplex
number $w = z_1 \ee + z_2 \eee$ as
\begin{equation}
w^{\dagger} := \bar{z}_1 \ee + \bar{z}_2 \eee ,
\label{2.4aa}
\end{equation}
where the bar denotes the usual complex conjugation.
Operation $w^{\dagger}$ was denoted by $w^{\dagger_3}$
in~\cite{Rochon3,Rochon2}, consistent with
the fact that at least two other types of conjugation
can be defined with bicomplex numbers.  Making use
of~(\ref{2.4cc}) we immediately see that
\begin{equation}
w \cdot w^{\dagger} = z_1 \bar{z}_1 \ee
+ z_2 \bar{z}_2 \eee .
\label{2.4bb}
\end{equation}
Furthermore, for any $s,t\in\T$,
\begin{equation}
(s+t)^{\dagger}=s^{\dagger}+t^{\dagger},\qquad
\left(s^{\dagger} \right)^{\dagger}=s,\qquad
(s\cdot t)^{\dagger}=s^{\dagger}\cdot t^{\dagger} .
\end{equation}

The real modulus $|w|$ of a bicomplex number $w$ can be
defined as
\begin{equation}
|w| := \sqrt{w_e^2 + w_{\ii}^2 + w_{\iii}^2 + w_j^2}
= \sqrt{(z_1 \bar{z}_1 + z_2 \bar{z}_2)/2} \, .
\label{2.8aa}
\end{equation}
This coincides with the Euclidean norm on $\R^4$.
Clearly, $|w| \ge 0$, with $|w| = 0$ if and only if $w=0$.
Moreover, one can show~\cite{Rochon1} that for any
$s,t\in\T$,
\begin{equation}
|s+t| \leq |s|+|t| , \qquad
|s\cdot t| \leq \sqrt{2} |s| \cdot |t| .
\label{2.16d}
\end{equation}
The modulus $|w|$ satisfies the first four properties
of the absolute value $N$ introduced in section~1,
but it fails to satisfy the fifth one.  The modulus can
be redefined so as to eliminate the $\sqrt{2}$
in~(\ref{2.16d}), but we are following Price's
conventions~\cite{Price}.

In the idempotent basis, any hyperbolic number
can be written as $x_1 \ee + x_2 \eee$, with
$x_1$ and $x_2$ in~$\R$.  We define the set~$\D^+$
of positive hyperbolic numbers as
\begin{equation}
\D^+:= \{ x_1 \ee + x_2 \eee ~|~ x_1, x_2 \in\R^+ \}.
\label{2.20}
\end{equation}
Clearly, $w \cdot w^{\dagger} \in \D^+$ for any
$w$ in $\T$.  We shall say that $w$ is in $\ee \R^+$
if $w = x_1 \ee$ and $x_1$ is in $\R^+$ (and similarly
with $\eee \R^+$).

\subsection{Modules, scalar product and
linear operators}
\label{Module}

By definition, a vector space is specified over a
field of numbers.  Bicomplex numbers make up a ring
rather than a field, and the structure analogous to
a vector space is then a \emph{module}.  For later
reference we define a $\T$-module $M$ as a set of
elements $\ket{\psi}$, $\ket{\phi}$, $\ket{\chi}$,
\ldots, endowed with operations of addition and
scalar multiplication, such that the following
always holds:
\begin{enumerate}
\item $\ket{\psi}+\ket{\phi}=\ket{\phi}+\ket{\psi}$;
\item $\left(\ket{\psi}+\ket{\phi}\right)+\ket{\chi}
=\ket{\psi}+\left(\ket{\phi}+\ket{\chi}\right)$;
\item there exists a $\ket{0}$ in $M$ such that
$\ket{0}+\ket{\psi}=\ket{\psi}$;
\item $0\cdot\ket{\psi}=\ket{0}$;
\item $1\cdot\ket{\psi}=\ket{\psi}$;
\item $s \cdot(\ket{\psi}+\ket{\phi})
= s \cdot\ket{\psi} + s \cdot\ket{\phi}$;
\item $\left(s+t \right)\cdot\ket{\psi}
= s \cdot\ket{\psi} + t \cdot\ket{\psi}$;
\item $(s t) \cdot\ket{\psi}
= s \cdot \left(t \cdot\ket{\psi}\right)$.
\end{enumerate}
Here $s, t \in \T$.  We have introduced Dirac's
notation for elements of $M$, which we shall call
\emph{kets} even though they are not genuine vectors.

A finite-dimensional \emph{free} $\T$-module is a $\T$-module
with a finite basis, \emph{i.e.} a finite set of linearly
independent elements that generate the module.  Explicitly,
$M$ is a finite-dimensional free $\T$-module if there exist
$n$ linearly independent kets $\ket{u_l}$ such that
any element $\ket{\psi}$ of $M$ can be written as
\begin{equation}
\ket{\psi} = \sum_{l=1}^n w_l\ket{u_l} ,\label{2.21}
\end{equation}
with $w_l \in \T$.  An important subset $V$ of $M$ is
the set of all kets for which all $w_l$ in~(\ref{2.21})
belong to $\C(\ii)$.  It was shown in~\cite{Rochon3}
that $V$ is a vector space over the complex numbers,
and that any $\ket{\psi} \in M$ can be decomposed
uniquely as
\begin{equation}
\ket{\psi} = \ee P_1 (\ket{\psi})
+ \eee P_2 (\ket{\psi}) ,
\label{2.21a}
\end{equation}
where $P_1$ and $P_2$ are projectors from $M$ to $V$.
Note that $V$ depends on the basis $\{\ket{u_l} \}$.
One can show that ket projectors
and idempotent-basis projectors (denoted with the
same symbol) satisfy the following, for $k=1, 2$:
\begin{equation}
\P{k}{s \ket{\psi} + t \ket{\phi}}
= \P{k}{s} \P{k}{\ket{\psi}}
+ \P{k}{t} \P{k}{\ket{\phi}} .
\end{equation}

It will be very useful to rewrite (\ref{2.10a})
and~(\ref{2.21a}) as
\begin{equation}
w = w_{\mathbf{1}} + w_{\mathbf{2}}, \qquad
\ket{\psi} = \ket{\psi}_{\mathbf{1}}
+ \ket{\psi}_{\mathbf{2}} ,
\label{2.21b}
\end{equation}
where
\begin{equation}
w_{\mathbf{1}} = \ee z_1, \quad w_{\mathbf{2}} = \eee z_2,
\quad \ket{\psi}_{\mathbf{1}} = \ee P_1 (\ket{\psi}) ,
\quad \ket{\psi}_{\mathbf{2}} = \eee P_2 (\ket{\psi}) .
\end{equation}
Henceforth bold indices
(like $\mathbf{1}$ and $\mathbf{2}$) will always
denote objects which include a
factor $\ee$ or $\eee$, and therefore satisfy an
equation like $w_{\mathbf{1}} = \ee w_{\mathbf{1}}$.

A \emph{bicomplex scalar product} maps two arbitrary kets
$\ket{\psi}$ and $\ket{\phi}$ into a bicomplex number
$(\ket{\psi}, \ket{\phi})$, so that the following
always holds ($s \in \T$):
\begin{enumerate}
\item[i.] $(\ket{\psi}, \ket{\phi} + \ket{\chi})
=(\ket{\psi}, \ket{\phi}) + (\ket{\psi}, \ket{\chi})$;
\item[ii.] $(\ket{\psi}, s \ket{\phi})
= s (\ket{\psi},\ket{\phi})$;
\item[iii.] $(\ket{\psi}, \ket{\phi})
= (\ket{\phi}, \ket{\psi})^{\dagger}$;
\item[iv.] $(\ket{\psi}, \ket{\psi})
=0~\Leftrightarrow~\ket{\psi}=0$.
\end{enumerate}
Property (iii) implies that $(\ket{\psi}, \ket{\psi})\in\D$.
In fact we will use in sections~3 and~4 the stronger
requirement that $(\ket{\psi}, \ket{\psi})\in\D^+$.
Writing property~(iii) with one of the other
types of conjugation mentioned after~(\ref{2.4aa})
would make $(\ket{\psi}, \ket{\psi})$ belong to
$\mathbb{C}(\ii)$ or $\mathbb{C}(\iii)$.

Properties (ii) and (iii) together imply that
$(s \ket{\psi}, \ket{\phi}) = s^{\dagger}
(\ket{\psi},\ket{\phi})$.  One easily shows that
\begin{equation}
(\ket{\psi}, \ket{\phi})
= (\ket{\psi}_{\mathbf{1}}, \ket{\phi}_{\mathbf{1}})
+ (\ket{\psi}_{\mathbf{2}}, \ket{\phi}_{\mathbf{2}}) .
\label{2.25}
\end{equation}
Note that
\begin{equation}
(\ket{\psi}_{\mathbf{1}}, \ket{\phi}_{\mathbf{1}})_{\mathbf{1}}
=(\ket{\psi}_{\mathbf{1}}, \ket{\phi}_{\mathbf{1}})
\quad \mbox{and} \quad
(\ket{\psi}_{\mathbf{2}}, \ket{\phi}_{\mathbf{2}})_{\mathbf{2}}
=(\ket{\psi}_{\mathbf{2}}, \ket{\phi}_{\mathbf{2}}).
\label{2.25a}
\end{equation}

A \emph{bicomplex linear operator} $A$ is a mapping
from $M$ to $M$ such that, for any $s, t \in \T$
and any $\ket{\psi}, \ket{\phi} \in M$
\begin{equation}
A (s \ket{\psi} + t \ket{\phi})
= s  A \ket{\psi} + t A \ket{\phi} .
\label{2.27}
\end{equation}
The bicomplex \emph{adjoint}
operator $A^*$ of $A$ is the operator
defined so that for any $\ket{\psi}, \ket{\phi} \in M$
\begin{equation}
(\ket{\psi}, A \ket{\phi})
= (A^* \ket{\psi}, \ket{\phi}) .\label{2.28}
\end{equation}
One can show that in finite-dimensional free $\T$-modules,
the adjoint always exists, is linear and satisfies
\begin{equation}
(A^*)^* = A, \quad (sA + tB)^*
= s^{\dagger}A^* + t^{\dagger} B^*, \quad (AB)^*=B^*A^* .
\end{equation}

A bicomplex linear operator $A$ can always be
written as $A = A_{\mathbf{1}} + A_{\mathbf{2}}$, with
$A_{\mathbf{1}} = \ee A$ and $A_{\mathbf{2}} = \eee A$.
Clearly,
\begin{equation}
A \ket{\psi} = A_{\mathbf{1}} \ket{\psi}_{\mathbf{1}}
+ A_{\mathbf{2}} \ket{\psi}_{\mathbf{2}} .
\end{equation}
We shall say that a ket $\ket{\psi}$ belongs to
the null cone if either $\ket{\psi}_{\mathbf{1}} = 0$ or
$\ket{\psi}_{\mathbf{2}} = 0$, and that a linear operator
$A$ belongs to the null cone if either $A_{\mathbf{1}} = 0$
or $A_{\mathbf{2}} = 0$.

A \emph{self-adjoint} operator is a linear operator $H$ such
that $H = H^*$.  From~(\ref{2.28}) one sees immediately
that $H$ is self-adjoint if and only if
\begin{equation}
(\ket{\psi}, H \ket{\phi}) = (H \ket{\psi}, \ket{\phi})
\end{equation}
for all $\ket{\psi}$ and $\ket{\phi}$ in $M$.

It was shown in~\cite{Rochon3} that
the eigenvalues of a self-adjoint operator acting
in a finite-dimensional free $\T$-module, associated
with eigenkets not in the null cone, are hyperbolic
numbers.  One can show quite straightforwardly that
two such eigenkets of such a self-adjoint operator, whose
eigenvalues differ by a quantity that is not in the
null cone, are orthogonal.  The proof of this statement
can be found as part of a detailed study of
finite-dimensional free $\T$-modules~\cite{Gervais}.

\section{The harmonic oscillator}
\label{BQHO}
\setcounter{equation}{0}
The harmonic oscillator is one of the most
widely discussed and widely applied problems
in standard quantum mechanics.  It is specified as
follows: Find the eigenvalues and eigenvectors
of a self-adjoint operator $H$ defined as
\begin{equation}
H=\frac{1}{2m}P^2+\frac{1}{2}m\omega^2X^2,
\label{3.1}
\end{equation}
where $m$ and $\omega$ are positive real
numbers and $X$ and $P$ are self-adjoint operators
satisfying the following commutation relation
(with $\ii$ the usual imaginary $\rmi$):
\begin{equation}
[X, P] = \ii \hbar I .\label{3.1a}
\end{equation}
The problem can be solved exactly either by
algebraic~\cite{Heisenberg,Dirac} or
differential~\cite{Eckart} methods.
In this section we shall
show that, viewed as an algebraic problem, the
standard quantum-mechanical harmonic oscillator
generalizes to bicomplex numbers.  One of the
advantages of the algebraic method is that the
uniqueness of the structure determined by the
assumptions is quite transparent.  In the process
we shall build explicitly an example of an
infinite-dimensional free $\T$-module.

\subsection{Definitions and assumptions}

To state and solve the problem of the bicomplex
quantum harmonic oscillator, we start with the
following assumptions:
\begin{itemize}
\item[a.] Three linear operators $X$, $P$ and $H$, related
by~(\ref{3.1}), act in a free $\T$-module $M$.
\item[b.] $X$, $P$ and $H$ are self-adjoint with
respect to a scalar product yet to be defined.
This means that
$(\ket{\psi}, H \ket{\phi}) = (H \ket{\psi}, \ket{\phi})$
for any $\ket{\psi}$ and $\ket{\phi}$ in $M$ on which
$H$ is defined, and similarly with $X$ and $P$.
\item[c.] The scalar product of a ket
with itself belongs to $\D^+$.
\item[d.] $[X, P] = \ii \hbar \xi I$, where $\xi\in\T$
is not in the null cone and $I$ is the identity
operator on $M$.
\item[e.] There is at least one
normalizable eigenket $\ket{E}$ of $H$ which is not
in the null cone and whose corresponding
eigenvalue $E$ is not in the null cone.
\item[f.] Eigenkets of $H$ that are not in the null
cone and that correspond to eigenvalues
whose difference is not in the null cone are orthogonal.
\end{itemize}
The consistency of these assumptions will be
verified explicitly once the full structure
has been obtained.  Note that not all of them
are expected to hold in an arbitrary
infinite-dimensional module, but they do in the
one we are going to construct.  The simplest extension
of the canonical commutation relations seems to
be embodied in~(d).  Note that (d) implies that
neither $X$ nor $P$ are in the null cone, for
if one of them were, $\xi$ would also belong
to~$\nc$. Assumption~(e) implies that $H$ is not
in the null cone, and this is necessary
to end up with a non-trivial generalization of the
standard quantum-mechanical case.

The self-adjointness of $X$ and $P$ implies that
the bicomplex number $\xi$ in~(d) is in fact
hyperbolic.  Indeed let $\ket{E}$ be the eigenket
of $H$ introduced in~(e).  By the
properties of the scalar product and definition of
self-adjointness,
{\begin{eqnarraz}
\ii \hbar \xi (\ket{E}, \ket{E})
&=& (\ket{E}, \ii \hbar \xi I \ket{E})
= (\ket{E}, XP \ket{E})- (\ket{E}, PX \ket{E}) \nonumber\\
&=& (X \ket{E}, P \ket{E})- (P \ket{E}, X \ket{E}) \nonumber\\
&=& (PX \ket{E}, \ket{E})- (XP \ket{E}, \ket{E}) \nonumber\\
&=& ( - \ii \hbar \xi I \ket{E}, \ket{E})
= \ii \hbar \xi^{\dagger} (\ket{E}, \ket{E}) .
\label{eq33}
\end{eqnarraz}}%
Since $\ket{E}$ is normalizable, $(\ket{E}, \ket{E})$
is not in the null cone, and
it immediately follows that $\xi = \xi^{\dagger}$.
That is, $\xi = \xi_{1} \ee + \xi_{2} \eee$,
with $\xi_{1}$ and $\xi_{2}$ real.

Is it possible to further restrict meaningful
values of $\xi$, for instance by a simple rescaling
of $X$ and $P$?  To answer this question, let us write
\begin{equation}
X=(\alpha_{1}\ee+\alpha_{2}\eee)X' , \qquad
P=(\beta_{1}\ee+\beta_{2}\eee)P' ,
\end{equation}
with non-zero
$\alpha_k$ and $\beta_k$ ($k=1, 2$).  For $X'$ and $P'$
to be self-adjoint, $\alpha_k$ and $\beta_k$ must be real.
Making use of~\eref{3.1} we find that
{\begin{eqnarraz}
H&=&\frac{1}{2m}(\beta_{1}^2\ee
+\beta_{2}^2\eee)(P')^2
+\frac{1}{2}m\omega^2(\alpha_{1}^2\ee
+\alpha_{2}^2\eee)(X')^2
\nonumber\\
&=&\frac{1}{2m'}(P')^2+\frac{1}{2}m'(\omega')^2(X')^2.
\label{3.2}
\end{eqnarraz}}%
For $m'$ and $\omega'$ to be positive real numbers,
$\alpha_{1}^2\ee+\alpha_{2}^2\eee$ and
$\beta_{1}^2\ee+\beta_{2}^2\eee$
must also belong to $\R^+$.  This entails
that $\alpha_{1}^2=\alpha_{2}^2$ and
$\beta_{1}^2=\beta_{2}^2$,
or equivalently
$\alpha_{1} = \pm \alpha_{2}$ and
$\beta_{1} = \pm \beta_{2}$.
Hence we can write
{\begin{eqnarraz}
\ii\hbar(\xi_{1} \ee+\xi_{2} \eee)I
&=& [X, P] \nonumber\\
&=& [(\alpha_{1}\ee+\alpha_{2}\eee)X',
(\beta_{1}\ee+\beta_{2}\eee)P']
\nonumber\\
&=&(\alpha_{1}\beta_{1}\ee+\alpha_{2}\beta_{2}\eee)[X',P'] .
\end{eqnarraz}}%
But this in turn implies that
\begin{equation}
[X',P']=\ii\hbar\left( \frac{\xi_{1}}
{\alpha_{1}\beta_{1}}\ee
+\frac{\xi_{2}}{\alpha_{2}
\beta_{2}}\eee \right)I
=\ii\hbar(\xi_{1} '\ee+\xi_{2} '\eee)I .
\label{3.3}
\end{equation}
This equation shows that $\alpha_{1}$,
$\alpha_{2}$, $\beta_{1}$
and $\beta_{2}$ can always be picked so that
$\xi_{1}'$ and
$\xi_{2}'$ are positive. Furthermore,
we can choose $\alpha_{1}$
and $\beta_{1}$ so as to make
$\xi_{1}'$ equal to~1.  But since
$|\alpha_{1} \beta_{1}|
= |\alpha_{2} \beta_{2}|$, we have no control
over the norm of $\xi_{2}'$.  The upshot is that we can
always write $H$ as in~\eref{3.1}, with the commutation
relation of $X$ and $P$ given by
\begin{equation}
[X,P] = \ii\hbar \xi I = \ii\hbar(\xi_{1}\ee+\xi_{2}\eee)I,
\qquad \xi_{1},\xi_{2}\in\R^+ . \label{3.5}
\end{equation}
We also have the freedom of setting either $\xi_{1} = 1$
or $\xi_{2} = 1$, but not both.

Just as in the case of the standard quantum
harmonic oscillator, we now introduce two operators
$A$ and $A^*$ as
{\begin{eqnarraz}
A&:=&\frac{1}{\sqrt{2m\hbar\omega}}(m\omega X + \ii P),
\label{3.7}\\
A^*&:=&\frac{1}{\sqrt{2m\hbar\omega}}(m\omega X - \ii P).
\label{3.8}
\end{eqnarraz}}%
Since $P$ is self-adjoint, one always has
$(-\ii P \ket{\psi}, \ket{\phi}) = (\ket{\psi}, \ii P\ket{\phi})$,
which means that the adjoint of $\ii P$ is $-\ii P$.  This
implies that, as the notation suggests,
$A^*$ is indeed the adjoint of $A$.  Equations~\eref{3.7}
and~\eref{3.8} can be inverted as
\begin{equation}
X = \sqrt{\frac{\hbar}{2m\omega}}(A+A^*),
\qquad P=-\ii\sqrt{\frac{\hbar m\omega}{2}}(A-A^*).
\label{3.9}
\end{equation}
The commutator of $A$ and $A^*$ is given by
\begin{equation}
[A,A^*] = \frac{1}{2m\hbar\omega}
\left\{ [\ii P, m\omega X]+[m\omega X, -\ii P] \right\}
= \xi I. \label{3.10}
\end{equation}
Substituting~\eref{3.9} in~\eref{3.1}, one easily
finds that
\begin{equation}
H=\hbar\omega\left(A^*A+\frac{\xi}{2}I\right)
= \hbar\omega\left(AA^*-\frac{\xi}{2}I\right).
\label{3.12}
\end{equation}
From~\eref{3.10} and~\eref{3.12}, the following commutation
relations are straightforwardly obtained:
\begin{equation}
[H,A] = - \hbar\omega\xi A, \qquad
[H,A^*] = \hbar\omega\xi A^* . \label{3.14}
\end{equation}
Our problem is therefore the following:  Find the eigenvalues
and eigenkets of the Hamiltonian~\eref{3.12}, subject to
the constraint~\eref{3.10} and, consequently,~\eref{3.14}.
\subsection{Eigenkets and eigenvalues of $H$}

From assumption~(e) we know that there is a normalizable
ket $\ket{E}$ such that
\begin{equation}
H\ket{E}=E\ket{E} .\label{3.15a}
\end{equation}
We can write
{\begin{eqnarraz}
H &=& H_{\mathbf{1}} + H_{\mathbf{2}},
\label{3.16a} \\
E &=& E_{\mathbf{1}} + E_{\mathbf{2}},
\label{3.16b} \\
\ket{E} &=& \ket{E}_{\mathbf{1}}
+ \ket{E}_{\mathbf{2}}, \label{3.16c}
\end{eqnarraz}}%
where $E_{\mathbf{1}} = \ee E$, etc.  Assumption~(e)
implies that none of the quantities
in~\eref{3.16a}--\eref{3.16c} vanishes.  Substitution
of these equations in~\eref{3.15a}
immediately yields
\begin{equation}
H_{\mathbf{1}} \ket{E}_{\mathbf{1}}
= E_{\mathbf{1}} \ket{E}_{\mathbf{1}} ,
\qquad H_{\mathbf{2}} \ket{E}_{\mathbf{2}}
= E_{\mathbf{2}} \ket{E}_{\mathbf{2}} .
\label{3.16}
\end{equation}

Following the treatment made in standard quantum
mechanics, we now apply operators $HA$ and $HA^*$ on
$\ket{E}$. Making use of~\eref{3.14} we get
{\begin{eqnarraz}
HA\ket{E} &=& (AH+[H,A])\ket{E}
= (E-\hbar\omega\xi )A\ket{E}, \label{3.17}\\
HA^*\ket{E} &=& (A^*H+[H,A^*])\ket{E}
= (E+\hbar\omega\xi)A^*\ket{E}.\label{3.18}
\end{eqnarraz}}%
We see that if $A\ket{E}$ does not vanish, it is an
eigenket of $H$ with eigenvalue $E-\hbar\omega\xi$.
Similarly, unless $A^*\ket{E}$ vanishes, it is an
eigenket of $H$ with eigenvalue $E+\hbar\omega\xi$.

Let $l$ be a positive integer.  We will show by
induction that unless $A^l\ket{E}$ vanishes, it is
an eigenket of $H$ with eigenvalue $E-l\hbar\omega\xi$.
We have just shown that this is true for $l=1$.  Let
it be true for $l-1$.  We have
{\begin{eqnarraz}
HA^{l} \ket{E} &=& HAA^{l-1}\ket{E}
=(AHA^{l-1} + [H,A]A^{l-1})\ket{E} \nonumber\\
&=& A(E-(l-1)\hbar\omega\xi)A^{l-1}\ket{E}
- \hbar\omega\xi AA^{l-1}\ket{E} \nonumber\\
&=& \left\{ E - l\hbar\omega\xi \right\} A^{l}\ket{E} ,
\label{3.19}
\end{eqnarraz}}%
which proves the claim.  Similarly, unless
$(A^*)^l \ket{E}$ vanishes, it is an eigenket
of $H$ with eigenvalue $E+l\hbar\omega\xi$,
that is,
\begin{equation}
H(A^*)^l\ket{E} = (E+l\hbar\omega\xi)(A^*)^l\ket{E} .
\label{3.20}
\end{equation}
Equations~\eref{3.19} and~\eref{3.20} separate
in the idempotent basis.  Multiplying them by $\e{k}$
and using the fact that $HA^l=H_{\mathbf{1}} A_{\mathbf{1}}^l
+ H_{\mathbf{2}} A_{\mathbf{2}}^l$, we easily find that
($k = 1, 2$)
{\begin{eqnarraz}
H_\mathbf{k}A_{\mathbf{k}}^l\ket{E}_{\mathbf{k}} 
&=& (E_{\mathbf{k}}-l\hbar\omega\xi_{\mathbf{k}})
A_{\mathbf{k}}^l\ket{E}_{\mathbf{k}} ,
\label{3.21}\\
H_{\mathbf{k}}(A_{\mathbf{k}}^*)^l\ket{E}_{\mathbf{k}}
&=& (E_{\mathbf{k}}+l\hbar\omega\xi_{\mathbf{k}})
(A_{\mathbf{k}}^*)^l\ket{E}_{\mathbf{k}} .
\label{3.22}
\end{eqnarraz}}%
Consistent with the bold notation, we have written
$\xi_{\mathbf{1}} = \ee \xi_1$ and
$\xi_{\mathbf{2}} = \eee \xi_2$.

We now prove the following lemma.
\begin{Lemma}
Let $\ket{\phi}$ be an eigenket of $H$ associated
with the (finite) eigenvalue $\lambda$. Then,
\begin{equation}
(A\ket{\phi},A\ket{\phi})
= \left\{ \frac{\lambda}{\hbar\omega}
- \frac{\xi}{2} \right\}(\ket{\phi},\ket{\phi})
\end{equation}
and
\begin{equation}
(A^*\ket{\phi},A^*\ket{\phi})
= \left\{ \frac{\lambda}{\hbar\omega}
+ \frac{\xi}{2} \right\}(\ket{\phi},\ket{\phi}).
\end{equation}
Proof.	\label{Lemma4.1}
\end{Lemma}
Making use of \eref{3.12} we have
{\begin{eqnarraz*}
(A\ket{\phi},A\ket{\phi})
&=& (\ket{\phi}, A^*A \ket{\phi})
= \left( \ket{\phi}, \left\{\frac{H}{\hbar\omega}
- \frac{\xi}{2}I\right\} \ket{\phi} \right) \\
&=& \left( \ket{\phi}, \left\{\frac{\lambda}{\hbar\omega}
- \frac{\xi}{2}\right\} \ket{\phi}  \right)
= \left\{\frac{\lambda}{\hbar\omega}-\frac{\xi}{2}\right\}
\left(\ket{\phi},\ket{\phi} \right) .
\end{eqnarraz*}}%
The proof of the second equality is similar.~$\Box$

Two important consequences of lemma~\ref{Lemma4.1}
are the following.  Firstly, whenever
$(\ket{\phi}, \ket{\phi})$ is finite,
so are $(A\ket{\phi},A\ket{\phi})$ and
$(A^*\ket{\phi},A^*\ket{\phi})$.  And secondly,
the lemma also holds when all quantities are
replaced by corresponding idempotent projections.
That is, for $k = 1, 2$,
\begin{equation}
(A_\mathbf{k} \ket{\phi}_\mathbf{k}, 
A_\mathbf{k} \ket{\phi}_\mathbf{k})
= \left\{ \frac{\lambda_\mathbf{k}}{\hbar\omega}
- \frac{\xi_\mathbf{k}}{2} \right\}(\ket{\phi}_\mathbf{k},
\ket{\phi}_\mathbf{k}) .
\label{3.20a}
\end{equation}

Let us now apply lemma~\ref{Lemma4.1} to the case
where $\ket{\phi}_\mathbf{k} = \ket{E}_\mathbf{k}$.  Since
$(\ket{E}, \ket{E})$ is in $\D^+$,
$(\ket{E}_\mathbf{k}, \ket{E}_\mathbf{k})$ is in 
$\mathbf{e_k}\R^+$ (and is non-zero).
But then~\eref{3.20a}
implies that $(A_{\mathbf{k}} \ket{E}_{\mathbf{k}}, A_{\mathbf{k}} \ket{E}_{\mathbf{k}})$ is in $\mathbf{e_k}\R^+$
only if $E_{\mathbf{k}} /\hbar \omega - \xi_{\mathbf{k}} /2$
is in $\mathbf{e_k}\R^+$.
Let us write~\eref{3.20a} for the case where
$\ket{\phi}_{\mathbf{k}} = A_{\mathbf{k}}^l 
\ket{E}_{\mathbf{k}}$.  Making use
of~\eref{3.21}, we find that
\begin{equation}
\left( A_{\mathbf{k}}^{l+1} \ket{E}_{\mathbf{k}},
A_{\mathbf{k}}^{l+1} \ket{E}_{\mathbf{k}} \right)
= \left\{ \frac{E_{\mathbf{k}}}{\hbar\omega}
- \left( l + \frac{1}{2} \right) \xi_{\mathbf{k}} \right\}
\left( A_{\mathbf{k}}^l \ket{E}_{\mathbf{k}},
A_{\mathbf{k}}^l \ket{E}_{\mathbf{k}} \right) .
\label{3.20b}
\end{equation}
Again, and assuming that $A_{\mathbf{k}}^l \ket{E}_{\mathbf{k}}$ doesn't
vanish, $(A_{\mathbf{k}}^{l+1} \ket{E}_{\mathbf{k}},
A_{\mathbf{k}}^{l+1} \ket{E}_{\mathbf{k}})$
is in $\mathbf{e_k}\R^+$ only if
$E_{\mathbf{k}} /\hbar \omega - (l+ 1/2) 
\xi_{\mathbf{k}}$ is in $\mathbf{e_k}\R^+$.

Clearly, however, this cannot go on forever.  Let
$l_k$ be the smallest positive integer for which
\begin{equation}
P_k \left(\frac{E_{\mathbf{k}}}{\hbar \omega}
- \left(l_k + \frac{1}{2} \right) \xi_{\mathbf{k}}\right) \le 0 .
\label{3.20c}
\end{equation}
If the equality holds in~\eref{3.20c}, then~\eref{3.20b}
implies that $A_{\mathbf{k}}^{l_{k}+1} \ket{E}_{\mathbf{k}} = 0$.  If the
inequality holds, the same conclusion follows since
otherwise the scalar product of a non-zero ket with
itself would be outside $\D^+$.  The upshot is that
\begin{equation}
A_{\mathbf{k}} \ket{\phi_0}_{\mathbf{k}} = 0 \quad \mbox{with}
\quad \ket{\phi_0}_{\mathbf{k}}
= A_{\mathbf{k}}^{l_k} \ket{E}_{\mathbf{k}} .
\end{equation}
Applying $H_{\mathbf{k}}$ obtained from the first part of~\eref{3.12}
on $\ket{\phi_0}_{\mathbf{k}}$, we get
\begin{equation}
H_{\mathbf{k}} \ket{\phi_0}_{\mathbf{k}} = \hbar \omega
\left(A_{\mathbf{k}}^* A_{\mathbf{k}} + \frac{1}{2} \xi_{\mathbf{k}} I \right)
\ket{\phi_0}_{\mathbf{k}} = \frac{1}{2} \hbar \omega \xi_{\mathbf{k}} \ket{\phi_0}_{\mathbf{k}} .
\end{equation}
That is, $\ket{\phi_0}_{\mathbf{k}}$ is an eigenket of $H_{\mathbf{k}}$
with eigenvalue $\hbar \omega \xi_{\mathbf{k}} /2$.

Making use of an argument similar to the one leading
to~\eref{3.22}, we can see that the ket
$(A_{\mathbf{k}}^*)^l \ket{\phi_0}_{\mathbf{k}}$ is an eigenket of $H_{\mathbf{k}}$
with eigenvalue $(l+1/2) \hbar \omega\xi_{\mathbf{k}}$. For
later convenience we define
\begin{equation}
\ket{\phi_l}_{\mathbf{1}} = (l! \xi_1^l)^{-1/2}
(A_{\mathbf{1}}^*)^l \ket{\phi_0}_{\mathbf{1}} , \qquad
\ket{\phi_l}_{\mathbf{2}} = (l! \xi_2^l)^{-1/2}
(A_{\mathbf{2}}^*)^l \ket{\phi_0}_{\mathbf{2}} .
\label{3.23a}
\end{equation}
Note that $\xi_1$ and $\xi_2$, being within an
inversion operator, cannot carry bold indices.
By the idempotent projection of the second part of
lemma~\ref{Lemma4.1}, $\ket{\phi_l}_{\mathbf{k}}$ does not
vanish for any $l$.  We have
therefore constructed two infinite sequences of
kets, each of which is a sequence of eigenkets
of an idempotent projection of $H$.

We now define
\begin{equation}
\ket{\phi_l} = \ket{\phi_l}_{\mathbf{1}}
+ \ket{\phi_l}_{\mathbf{2}} .
\label{3.23b}
\end{equation}
It is easy to check that $\ket{\phi_l}$ is an eigenket
of $H$ with eigenvalue $(l+1/2) \hbar \omega \xi$.
By assumption~(f), $\ket{\phi_l}$ and
$\ket{\phi_{l'}}$ are orthogonal if $l \neq l'$.

\subsection{Infinite-dimensional free $\T$-module}

Let $M$ be the collection of all finite linear
combinations of kets $\ket{\phi_l}$, with bicomplex
coefficients.  That is,
\begin{equation}
M := \left\{\sum_{l}{w_l\ket{\phi_l}}
~|~w_l \in\T\right\} . \label{3.29}
\end{equation}
It is understood that adding terms with zero
coefficients doesn't yield a new ket.
Let us define the addition of two elements of~$M$
and the multiplication of an element of~$M$ by a
bicomplex number in the obvious way.  Furthermore
let us write $\ket{0} = 0 \cdot \ket{\phi_0}$.
It is then easy to check that the eight defining
properties of a $\T$-module stated in section~2.2
are satisfied.  $M$ is therefore a $\T$-module.
Clearly, $H$ is defined everywhere on~$M$.

If the coefficients $w_l$ in~\eref{3.29} are
restricted to elements of~$\mathbb{C}(\ii)$,
the resulting set $V$ is a vector space
over~$\mathbb{C}(\ii)$.  It is the analog of
the vector space introduced before~\eref{2.21a},
which was used in~\cite{Rochon3} to define the
projection $P_k$ and prove a number of results
on finite-dimensional modules.

The scalar product of elements of~$M$ has
hitherto been specified only partially, in
particular by requiring that $\ket{\phi_l}$ and
$\ket{\phi_{l'}}$ be orthogonal if $l \neq l'$.
We now set
\begin{equation}
(\ket{\phi_0}, \ket{\phi_0}) = 1 .
\label{3.30a}
\end{equation}
Equation~\eref{3.23a} implies that
{\begin{eqnarraz}
\ket{\phi_{l+1}} &=& \ket{\phi_{l+1}}_{\mathbf{1}}
+ \ket{\phi_{l+1}}_{\mathbf{2}} 
= \ee \ket{\phi_{l+1}}_{\mathbf{1}}
+ \eee \ket{\phi_{l+1}}_{\mathbf{2}} \nonumber \\
&=& \frac{\ee}{\sqrt{(l+1) \xi_1}}
A_{\mathbf{1}}^* \ket{\phi_l}_{\mathbf{1}} 
+ \frac{\eee}{\sqrt{(l+1) \xi_2}}
A_{\mathbf{2}}^* \ket{\phi_l}_{\mathbf{2}} \nonumber \\
&=& \frac{1}{\sqrt{(l+1) \xi}}
A^* \ket{\phi_l} .
\label{3.30b} 
\end{eqnarraz}}%
Letting $A$ act on both sides of~\eref{3.30b}
and making use of~\eref{3.12}, we find that
{\begin{eqnarraz}
A \ket{\phi_{l+1}} 
&=& \frac{1}{\sqrt{(l+1) \xi}} A A^* \ket{\phi_l} 
= \frac{1}{\sqrt{(l+1) \xi}} \left\{ 
\frac{H}{\hbar \omega} + \frac{\xi}{2} I
\right\} \ket{\phi_l} \nonumber \\ 
&=& \sqrt{(l+1) \xi} \, \ket{\phi_l} .
\label{3.30cc}
\end{eqnarraz}}%
Equations~\eref{3.30b} and~\eref{3.30cc} imply that
$A$ and $A^*$, and therefore $X$ and $P$, are defined
everywhere on~$M$.  This means that all quantities
involved, for instance, in~\eref{eq33} are well-defined.

From~\eref{3.30b} and the second part of
lemma~\ref{Lemma4.1} we get
\begin{equation}
(\ket{\phi_{l+1}}, \ket{\phi_{l+1}})
= \frac{1}{(l+1) \xi}
(A^* \ket{\phi_{l}}, A^* \ket{\phi_{l}})
= (\ket{\phi_{l}}, \ket{\phi_{l}}) .
\end{equation}
Owing to~\eref{3.30a}, the solution of this
recurrence equation is
\begin{equation}
(\ket{\phi_l}, \ket{\phi_l}) = 1, \qquad l = 0, 1, 2, \ldots
\label{3.30c}
\end{equation}

We now fully specify the scalar product of two
arbitrary elements $\ket{\psi}$ and $\ket{\chi}$
of~$M$ as follows.  Let
\begin{equation}
\ket{\psi} = \sum_{l} w_l \ket{\phi_l} , \qquad
\ket{\chi} = \sum_{l} v_l \ket{\phi_l} .
\label{3.31a}
\end{equation}
The two sums are finite.  Without loss of
generality, we can let them run over the same
set of indices.  Indeed this simply amounts
to possibly adding terms with zero coefficients
in either or both sums.  With this we define the scalar
product as
\begin{equation}
(\ket{\psi}, \ket{\chi}) 
:= \sum_{l} w_l^{\dagger} v_l (\ket{\phi_l}, \ket{\phi_l}) 
= \sum_{l} w_l^{\dagger} v_l .
\label{3.31b}
\end{equation}
With this specification, it is easy to check
that the four defining properties of a scalar
product stated in section~2.2 are satisfied.
Note that the right-hand side of~\eref{3.31b}
is always finite.

Clearly, the kets $\ket{\phi_l}$ generate $M$.
To show that they are linearly independent, we
assume that $\ket{\psi}$ defined in~\eref{3.31a}
vanishes.  Letting $m$ be one of the $l$ indices,
we have
\begin{equation}
0 = (\ket{\phi_m}, \ket{\psi}) =
\sum_{l} w_l \delta_{ml} = w_m .
\label{3.31c}
\end{equation}
Hence $w_m = 0$ for all $m$, and the linear
independence follows.  This shows that~$M$ is
an infinite-dimensional free $\T$-module.

There remains to check the six assumptions
made at the beginning of section~3.1.
Assumption~(a) is obvious, the action of $X$ and
$P$ on $M$~being most easily obtained through
the action of $A$ and $A^*$.  Similarly with~(b),
the self-adjointness of $X$ and $P$ follows from
the easily verifiable fact that $A^*$ is the
adjoint of $A$ on the whole of~$M$.  Assumption~(c)
is an immediate consequence of definition~\eref{3.31b}.
Assumption~(d) follows from the commutation
relation $[A, A^*] = \xi I$.  This one is easily
checked when acting on eigenkets of~$H$ and therefore,
by linearity, it holds on any ket.
Assumption~(e) is satisfied by any ket $\ket{\phi_l}$.
There only remains to check assumption~(f),
which is a little more tricky.

Let $\ket{\psi}$ defined in~\eref{3.31a} be an
eigenket of $H$ with eigenvalue~$\lambda$.
This means that
\begin{equation}
H \sum_l w_l \ket{\phi_l} = \lambda \sum_l w_l \ket{\phi_l}
\end{equation}
which, owing to the linear independence of the
$\ket{\phi_l}$, reduces to
\begin{equation}
\left(l + \frac{1}{2} \right) \hbar \omega \xi w_l 
= \lambda w_l .
\end{equation}
In the idempotent basis this becomes ($k = 1, 2$)
\begin{equation}
\left(l + \frac{1}{2} \right) \hbar \omega 
\xi_{\mathbf{k}} w_{l \mathbf{k}} 
= \lambda_{\mathbf{k}} w_{l \mathbf{k}} .
\end{equation}
Let $\lambda_{\mathbf{1}} \neq 0$. Since
$\xi_{\mathbf{1}}$ does not vanish, at most
one coefficient $w_{l\mathbf{1}}$ does not vanish,
for otherwise $\lambda_{\mathbf{1}}$ would satisfy
two incompatible equations.  If
$\lambda_{\mathbf{1}} = 0$, all $w_{l\mathbf{1}}$ vanish.
A similar argument holds for $\mathbf{2}$.  Hence
the eigenket of $H$ has the form
\begin{equation}
\ket{\phi} = w_{l\mathbf{1}} \ket{\phi_l}_{\mathbf{1}}
+ w_{l'\mathbf{2}} \ket{\phi_{l'}}_{\mathbf{2}} ,
\label{3.32a}
\end{equation}
with one of the coefficients vanishing if the
corresponding $\lambda_{\mathbf{k}}$ vanishes.  If both
$\lambda_{\mathbf{k}}$ vanish, all 
$w_{l \mathbf{k}} = 0$ and there is no
eigenket.  The upshot is that~\eref{3.32a} represents
the most general eigenket of $H$. Its associated
eigenvalue $\lambda$ is
\begin{equation}
\lambda = \hbar \omega \left\{ \left(
l + \frac{1}{2} \right) \xi_1 \ee
+ \left( l' + \frac{1}{2} \right) \xi_2 \eee \right\} .
\end{equation}

It is now a simple matter to check that
assumption~(f) is satisfied.  Note that the
restriction on the difference of eigenvalues
cannot be dispensed with.  Indeed the two kets
\begin{equation}
\ket{\phi} = \ket{\phi_1}_{\mathbf{1}}
+ \ket{\phi_{2}}_{\mathbf{2}} , \qquad
\ket{\phi'} = \ket{\phi_1}_{\mathbf{1}}
+ \ket{\phi_{3}}_{\mathbf{2}}
\end{equation}
are examples of eigenkets that correspond to
different eigenvalues whose difference is in the
null cone.  Clearly, they are not orthogonal.

\section{Harmonic oscillator wave functions}
\setcounter{equation}{0}
\subsection{Bicomplex function space}

Let $n$ be a non-negative integer and let $\alpha$
be a positive real number.  Consider the following
function of a real variable $x$:
\begin{equation}
f_{n,\alpha} (x) := x^n \exp (-\alpha x^2) .
\label{4ff}
\end{equation}
Let $S$ be the set of all finite linear combinations
of functions $f_{n,\alpha} (x)$, with complex
coefficients.  Furthermore, let a bicomplex function
$u(x)$ be defined as
\begin{equation}
u(x) = \ee u_{1} (x) + \eee u_{2} (x) ,
\label{4fu}
\end{equation}
where $u_{1}$ and $u_{2}$ are both in~$S$.  It is then
easy to check that the set of all functions $u(x)$
is a $\T$-module, which we shall denote by $M_S$.

Let $u(x)$ and $v(x)$ both belong to $M_S$.
We define a mapping $(u, v)$ of this pair of
functions into~$\D^+$ as follows:
\begin{equation}
\left(u, v \right)
:= \int_{-\infty}^{\infty} u^{\dagger}(x) v(x) \rmd x 
= \int_{-\infty}^{\infty} \left[ \ee \bar{u}_{1} (x) 
v_{1} (x) + \eee \bar{u}_{2} (x) v_{2} (x) \right] \rmd x .
\label{3.36}
\end{equation}
It is not hard to see that~\eref{3.36}
is always finite and satisfies all the
properties of a bicomplex scalar product.

Let $\xi \in \D^+$.  We define two operators $X$ and $P$
that act on elements of $M_S$ as follows:
\begin{equation}
X \{u(x)\} := x u(x) , \qquad 
P \{u(x)\} := -\ii \hbar \xi \frac{\rmd u(x)}{\rmd x} .
\label{3.35}
\end{equation}
It is not difficult to show that $[X, P] = \ii \hbar \xi I$.
Note that
{\begin{eqnarraz}
X \{f_{n,\alpha} (x)\} &=& f_{n+1,\alpha} (x) ,
\label{3.35a} \\ 
P \{f_{n,\alpha} (x)\} &=& - \ii \hbar \xi 
\left\{ n f_{n-1, \alpha} (x) - 2 \alpha f_{n+1,\alpha} (x) \right\}.
\label{3.35b}
\end{eqnarraz}}%
From this we conclude that the action of $X$ and $P$ on
elements of $M_S$ always yields elements of $M_S$ (the
function $f_{-1, \alpha}$, if any, coming with a
vanishing coefficient).  That is,
$X$ and $P$ are defined on all $M_S$.

One can easily check that $(Xu, v) = (u, Xv)$,
so that $X$ is self-adjoint.  The self-adjointness
of $P$ can be proved as
{\begin{eqnarraz*}
&& \left( Pu, v \right) - \left( u, Pv \right) \\
&& \qquad = \int_{-\infty}^{\infty} 
\left(-\ii \hbar \xi \frac{\rmd u(x)}
{\rmd x}\right)^{\dagger} v(x) \rmd x 
- \int_{-\infty}^\infty u^{\dagger} (x) 
\left(-\ii \hbar \xi \frac{\rmd v(x)}{\rmd x}\right) \rmd x \\
&& \qquad = \ii \hbar \xi \left\{ \int_{-\infty}^{\infty}
\frac{\rmd \left[u^{\dagger}(x) v(x) \right]}
{\rmd x} \rmd x \right\} \\
&& \qquad = \ii \hbar \xi \left[u^{\dagger}(x) v(x) 
\right]_{-\infty}^{\infty}  = 0 .
\label{3.38}
\end{eqnarraz*}}%
The final equality comes from the fact that
$u$ and $v$, involving finite sums of functions
$f_{n,\alpha} (x)$, vanish at infinity.

\subsection{Eigenfunctions of $H$}

Let $H$ be defined as in~\eref{3.1}, with $X$ and $P$
specified as in~\eref{3.35}.  The eigenvalue equation
for $H$ is then given by
\begin{equation}
H u(x) = - \frac{\hbar^2 \xi^2}{2m}
\frac{\rmd^2 u(x)}{\rmd x^2} 
+ \frac{1}{2} m \omega^2 x^2 u(x) = E u(x) .
\label{3.36a}
\end{equation}
In the idempotent basis this separates into the
following two equations ($k = 1, 2$):
\begin{equation}
- \frac{\hbar^2 \xi_k^2}{2m} \frac{\rmd^2 u_k(x)}{\rmd x^2} 
+ \frac{1}{2} m \omega^2 x^2 u_k(x) = E_k u_k(x) .
\label{3.36b}
\end{equation}
Each of these equations is essentially the
eigenvalue equation for the Hamiltonian of
the standard quantum harmonic oscillator.  The
only difference is that $\hbar$ is replaced
by $\hbar \xi_k$.

The eigenfunction associated with the lowest
eigenvalue of~\eref{3.36b} is given by
\begin{equation}
\phi_{0k} (x) 
= \left( \frac{m\omega}{\pi\hbar\xi_k} \right)^{1/4}
\expm{- \frac{m\omega}{2\hbar\xi_k}x^2}.
\label{3.45}
\end{equation}
The corresponding eigenfunction of $H$ is therefore
given by
{\begin{eqnarraz}
\phi_0 (x) &=& \ee \phi_{01} (x)
+ \eee \phi_{02} (x) \nonumber\\
&=& \ee\left( \frac{m\omega}{\pi\hbar\xi_{1}} \right)^{1/4}
\expm{-\frac{m\omega}{2\hbar\xi_{1}}x^2}
+ \eee\left( \frac{m\omega}{\pi\hbar\xi_{2}} \right)^{1/4}
\expm{-\frac{m\omega}{2\hbar\xi_{2}}x^2} \nonumber\\
&=& \left( \frac{m\omega}{\pi\hbar} \right)^{1/4}
\left(\frac{\ee}{\xi_{1}^{1/4}}
+ \frac{\eee}{\xi_{2}^{1/4}}\right)
\left\{ \ee \exp \left[ -\frac{m\omega}{2\hbar\xi_{1}} x^2 \right]
+ \eee \exp \left[-\frac{m\omega}{2\hbar\xi_{2}}x^2
\right] \right\}. \qquad
\label{3.46}
\end{eqnarraz}}%
It can be shown~\cite{Price} that for any
bicomplex number $w = z_{1} \ee + z_{2} \eee$,
\begin{equation}
\expm{w} = \ee \expm{z_{1}} + \eee \expm{z_{2}} . 
\label{3.46a}
\end{equation}
This holds also for any polynomial function
$Q(w)$, that is,
\begin{equation}
Q(z_{1} \ee + z_{2} \eee) 
= \ee Q(z_{1}) + \eee Q(z_{2}) . 
\label{3.46b}
\end{equation}
Moreover, if $\xi = \xi_{1} \ee + \xi_{2} \eee$
with $\xi_{1}$ and $\xi_{2}$ positive, we have
\begin{equation}
\frac{1}{\xi^{1/4}} = \frac{\ee}{\xi_{1}^{1/4}}
+ \frac{\eee}{\xi_{2}^{1/4}} . 
\label{3.46c}
\end{equation}
Substituting~\eref{3.46a} and~\eref{3.46c}
in~\eref{3.46}, we get
\begin{equation}
\phi_0 (x) = \left( \frac{m\omega}{\pi\hbar\xi} \right)^{1/4}
\expm{-\frac{m\omega}{2\hbar\xi}x^2} . 
\label{3.47}
\end{equation}
From the normalization of $\phi_{01}$ and
$\phi_{02}$, we find that
{\begin{eqnarraz}
\left(\phi_0 , \phi_0 \right)
&=& \int_{-\infty}^\infty \left[ \ee \bar{\phi}_{01} (x)
\phi_{01} (x) + \eee \bar{\phi}_{02} (x) 
\phi_{02} (x) \right] \rmd x \nonumber \\
&=& \ee + \eee = 1 .
\label{3.48}
\end{eqnarraz}}%

The eigenfunction associated with the $l$th
eigenvalue of~\eref{3.36b} is given by~\cite{Marchildon} 
\begin{equation}
\phi_{lk} (x) = \left[ \sqrt{\frac{m\omega}{\pi\hbar\xi_k}}
\frac{1}{2^l l!} \right]^{1/2}
\rme^{- \theta_k^2/2} H_l(\theta_k) ,
\label{3.49}
\end{equation}
where
\begin{equation}
\theta_k = \sqrt{\frac{m\omega}{\hbar\xi_k}} x 
\label{3.50}
\end{equation}
and $H_l(\theta_k)$ is the Hermite polynomial of order $l$.
Just as in~\eref{3.23b} we now define
\begin{equation}
\phi_l (x) = \ee \phi_{l1} (x)
+ \eee \phi_{l2} (x). 
\label{3.50a}
\end{equation}
We therefore obtain
{\begin{eqnarraz}
\phi_l (x)
&=& \ee\left[ \sqrt{\frac{m\omega}{\pi\hbar\xi_{1}}}
\frac{1}{2^ll!} \right]^{1/2}
\rme^{-\theta_1^2/2} H_l(\theta_1)
+ \eee\left[ \sqrt{\frac{m\omega}{\pi\hbar\xi_{2}}}
\frac{1}{2^ll!} \right]^{1/2}
\rme^{-\theta_2^2/2} H_l(\theta_2) \nonumber\\
&=& \left\{ \ee\left[ \sqrt{\frac{m\omega}{\pi\hbar\xi_{1}}}
\frac{1}{2^ll!} \right]^{1/2}
+ \eee\left[ \sqrt{\frac{m\omega}{\pi\hbar\xi_{2}}}
\frac{1}{2^ll!} \right]^{1/2} \right\} \nonumber\\
&& \qquad \cdot
\left\{\ee \rme^{-\theta_1^2/2} 
+ \eee \rme^{-\theta_2^2/2} \right\}
\left\{\ee H_l(\theta_1) + \eee H_l(\theta_2)\right\} .
\label{3.51}
\end{eqnarraz}}%
Letting $\theta := \ee \theta_{1} +
\eee \theta_{2}$ and making use
of~\eref{3.46a}--\eref{3.46c}, we finally obtain 
\begin{equation}
\phi_l (x) = \left[ \sqrt{\frac{m\omega}{\pi\hbar\xi}}
\frac{1}{2^l l!} \right]^{1/2} \rme^{-\theta^2/2}
H_l(\theta) ,
\label{3.53}
\end{equation}
where
\begin{equation}
H_l(\theta) := \ee H_l(\theta_{1})
+ \eee H_l(\theta_{2})
\label{4herm} 
\end{equation}
is a hyperbolic Hermite polynomial of order $l$.

Equation~\eref{3.53} is one of the central results
of this paper.  It expresses normalized eigenfunctions
of the bicomplex harmonic oscillator Hamiltonian purely
in terms of hyperbolic constants and functions, with no
reference to a particular representation like $\{ \e{k} \}$.
Indeed $\xi$ can be viewed as a $\D^+$ constant, $\theta$ is
equal to $\sqrt{m \omega / \hbar \xi} \, x$ and $H_l(\theta)$
is just the Hermite polynomial in $\theta$.

Let $\tilde{M}$ be the collection of all finite linear
combinations of bicomplex functions $\phi_l (x)$,
with bicomplex coefficients.  That is,
\begin{equation}
\tilde{M} := \left\{\sum_{l} w_l \phi_l (x)
~|~w_l \in\T\right\} . \label{3.53a}
\end{equation}
It is easy to see that $\phi_l (x)$ is a
function like $u(x)$ defined in~\eref{4fu}.  Thus
$\tilde{M}$ is a submodule of the module $M_S$ defined
earlier in terms of functions $u(x)$.  Moreover,
$\tilde{M}$ is isomorphic to the module $M$ defined in
section~3.3.

In section~3.3, the most general eigenket of $H$ was
written as in~\eref{3.32a}.  The corresponding
eigenfunction has the form
\begin{equation}
\phi (x) = \ee w_{l1} \phi_{l1} (x)
+ \eee w_{l'2} \phi_{l' 2} ,
\label{3.32b}
\end{equation}
with $w_{l1}$ and $w_{l'2}$ in $\mathbb{C}(\ii)$.
The eigenfunction can be written explicitly as
\begin{equation}
\phi (x) = \left[ \frac{m\omega}{\pi\hbar} \right]^{1/4}
\left\{\ee \frac{w_{l1} \rme^{-\theta_{1}^2/2}}
{\sqrt{2^l l! \sqrt{\xi_{1}}}} H_l(\theta_{1})
+ \eee \frac{w_{l'2} \rme^{-\theta_{2}^2/2}}
{\sqrt{2^{l'} (l')! \sqrt{\xi_{2}}}} H_{l'}
(\theta_{2}) \right\} .
\label{3.54}
\end{equation}
The function $\phi$ is normalized, \emph{i.e.} $(\phi, \phi) = 1$,
if
\begin{equation}
|w_{l1}|^2 \ee + |w_{l'2}|^2 \eee = 1 .
\end{equation}
This means that $|w_{l1}| = 1 = |w_{l'2}|$.  From
the properties of real Hermite polynomials, one can
show that two functions $\phi (x)$ associated with
eigenvalues whose difference is not in the null
cone, are orthogonal.

The function $\phi (x)$ can also be expressed
in terms of the hyperbolic units 1 and $\jj$
instead of $\ee$ and $\eee$.  Letting
$w_{l1} = 1 = w_{l'2}$, we get
{\begin{eqnarraz}
\phi (x) &=& \left[ \frac{m\omega}{\pi\hbar} \right]^{1/4}
\frac{1}{2} \left\{
\left[ \frac{\rme^{-\theta_{1}^2/2}}
{\sqrt{2^l l! \sqrt{\xi_{1}}}} H_l(\theta_{1})
+ \frac{\rme^{-\theta_{2}^2/2}}
{\sqrt{2^{l'} (l')! \sqrt{\xi_{2}}}} H_{l'}
(\theta_{2}) \right] \right. \nonumber\\
&& \qquad + \jj \left.
\left[ \frac{\rme^{-\theta_{1}^2/2}}
{\sqrt{2^l l! \sqrt{\xi_{1}}}} H_l(\theta_{1})
- \frac{\rme^{-\theta_{2}^2/2}}
{\sqrt{2^{l'} (l')! \sqrt{\xi_{2}}}} H_{l'}
(\theta_{2}) \right] \right\} .
\label{3.55}
\end{eqnarraz}}%
%

\section{Outlook and conclusion}
\setcounter{equation}{0}
Bicomplex numbers include complex numbers as a subset,
just like hyperbolic numbers include real numbers.
It should come as no surprise then that the eigenkets
and eigenfunctions of the standard harmonic oscillator
Hamiltonian can be recovered from the bicomplex ones.
This, in fact, can be done in three different ways,
as can be seen for instance from~\eref{3.32a}.
First we set $\xi_1 = 1 = \xi_2$.
We then let either (i) $w_{l\mathbf{1}} = \ee$ and
$w_{l'\mathbf{2}} = 0$, or (ii) $w_{l\mathbf{1}} = 0$ and
$w_{l'\mathbf{2}} = \eee$, or finally (iii) $l = l'$,
$w_{l\mathbf{1}} = \ee$ and $w_{l'\mathbf{2}} = \eee$.
In each case the eigenkets make up a structure
isomorphic to the standard harmonic oscillator
eigenvectors.

The usual Hermite polynomials can similarly be
recovered from the bicomplex ones.  Looking
at~\eref{4herm}, we can see that both $P_1 (H_l (\theta))$
and $P_2 (H_l (\theta))$ are real-valued Hermite
polynomials.  So is $H_l (\theta)$ itself, in the
special case where $\theta_1 = \theta_2 = \theta$.

In the module~$\tilde{M}$ defined
in~\eref{3.53a}, the coefficients $w_l$ are bicomplex
numbers.  If they are restricted to elements of
$\mathbb{C}(\ii)$, then the set
of linear combinations makes up a vector space $\tilde{V}$,
isomorphic to the space $V$ defined after~\eref{3.29}.
The space~$\tilde{V}$ is not restricted to standard
Hermite polynomials but contains all the
hyperbolic ones.

We should note that the infinite-dimensional
modules~$M$, $\tilde{M}$ and~$M_S$ that we have
introduced have been defined, so to speak, in a
minimal way.  Indeed they do contain all eigenkets
or eigenfunctions of the bicomplex harmonic oscillator
Hamiltonian, but they only involve finite
linear combinations of elements.  This was done
purposely, for in this way we are able to avoid
convergence problems or other delicate issues
involved in complete spaces like $L^2(\R)$.  Our
modules do not have a property of completeness.
Indeed Cauchy sequences of elements of $\tilde{M}$,
for instance, do not in general converge to
elements of~$\tilde{M}$.

We believe, however, that all our modules can be
extended to complete ones, that is, to what can be
called infinite-dimensional bicomplex Hilbert spaces.
To motivate this, let $\ket{\phi}$ be an arbitrary ket
in $M$.  We can define the norm of $\ket{\phi}$ as
the modulus of the square root of the scalar product of
$\ket{\phi}$ with itself:
\begin{equation}
\mbox{Norm} \, (\ket{\phi}) :=
\left| \sqrt{(\ket{\phi}, \ket{\phi})} \right| .
\end{equation}
One can show that up to a positive factor, this norm 
coincides with the square root of the sum of the squares
of the natural norms of $P_1(\ket{\phi})$ and
$P_2(\ket{\phi})$.  If we define a Cauchy sequence
of elements of $M$ as one for which
$\mbox{Norm} \, (\ket{\phi^m} - \ket{\phi^n}) \rightarrow 0$
as $m, n \rightarrow \infty$, then one can show that
a sequence of elements of $M$ is a Cauchy sequence
if and only if both its projections $P_1$ and $P_2$ are
Cauchy sequences.  This suggests that a bicomplex
Hilbert space can be built from the two separate
Hilbert spaces of its projections.  This is presently
being investigated.

We have shown that the standard
quantum-mechanical harmonic oscillator can
be generalized to bicomplex numbers.  We have not
suggested any specific physical interpretation of
the bicomplex eigenkets or eigenfunctions.  Indeed
much work has to be done to assess the extent to which
the postulates of standard quantum mechanics can
be extended to the bicomplex number system.  It
appears, however, that other
quantum-mechanical problems can be generalized
to bicomplex numbers.  Here we specifically
have in mind the $r^{-1}$ potential, which also
lends itself both to algebraic and differential equation
treatments~\cite{Marchildon}.  Presumably, real Laguerre
polynomials and complex spherical harmonics can be
generalized, respectively, to hyperbolic and bicomplex
ones.  This is also an area worth being investigated.

\section*{Acknowledgments}
DR is grateful to the Natural Sciences and
Engineering Research Council of Canada for financial
support.  RGL would like to thank the Qu\'{e}bec
FQRNT Fund for the award of a postgraduate
scholarship.

\end{document}